\documentclass[prb,aps,twocolumn,superscriptaddress,floatfix,showpacs]{revtex4}
\usepackage{graphicx}
\usepackage{amssymb}
\usepackage{amsmath}
\usepackage{xspace}
\usepackage{soul}
\begin{document}

\title{Metallic ferromagnetism-insulating charge order transition in doped manganites}
\author{Van-Nham Phan}
\affiliation{Institute of Research and Development, Duy Tan University, K7/25 Quang Trung, Danang, Vietnam}
\author{Quoc-Huy Ninh}
\affiliation{Natural Faculty of Science, Langson College of Education, 9-Deo Giang, Chilang, Langson, Vietnam}
\author{Minh-Tien Tran}
\affiliation{Institute of Research and Development, Duy Tan University, K7/25 Quang Trung, Danang, Vietnam}
\affiliation{Institute of Physics, Vietnam Academy of Science and Technology, 10 Daotan, Hanoi, Vietnam}

\begin{abstract}
We show that an interplay of double exchange and impurity randomness can explain the competition between metal-ferromagnetic and insulating charge ordered states in doped manganites. The double exchange is simplified in the Ising type, whereas the randomness is modeled by the Falicov-Kimball binary distribution. The combined model is considered in a framework of dynamical mean-field theory. Using the Kubo-Greenwood formalism, the transport coefficients are explicitly expressed in terms of single particle spectral functions. Dividing the system into two sublattices we have pointed out a direct calculation to the checkerboard charge order parameter and the magnetizations. Numerical results show us that the checkerboard charge order can settle inside the ferromagnetic state at low temperature. An insulator-metal transition is also found at the point of the checkerboard charge order-ferromagnetic transition.
\end{abstract}
\date{\today}
\maketitle
\section{Introduction}
Observations of the complex phase structures in doped manganites $T_{1-x}D_{x}$MnO$_3$ ($T$=trivalent rare earth, $D$=divalent alkaline) have stimulated enormous attention, recently.~\cite{ci01,ci02,ci03} In a wide range of doping $x$ and temperature, the competition of various degrees of freedoms leads to a very rich phase diagram involving spin, charge, and orbital orders.~\cite{ci1,ci2,ci3,ci4} Particularly, in some regimes of the phase diagram, an interplay between the charge and the spin orders has triggered various anomalies of the transport properties observed in experiments.~\cite{ci1,ci2,ci3,ci4} In doped manganites with perovskite structure, the Mn five-fold $3d$-levels are split into triply degenerate $t_{2g}$ and higher energy doubly degenerate $e_g$ levels. The $t_{2g}$ electrons are usually localized. Meanwhile, the $e_g$ electrons are able to hop between Mn sites, building the conducting band. The itinerant electrons and local spins are correlated by the double-exchange (DE) mechanism, in which two motions involving an itinerant electron moving from an oxygen atom to Mn$^{4+}$ ions and another from Mn$^{3+}$ to the oxygen atom happen simultaneously.~\cite{ci5} The main feature of DE is interactive cooperation between the ferromagnetic (FM) ordering of the local spin and the motion of the itinerant electrons. The DE model has successfully described some magnetic properties of manganites, in such a way that it provides a well-established starting point toward understanding the complex phase orders.

However, considering the DE model alone, only the ferromagnetic state stabilizes at low temperature and in the whole temperature range, one finds only the metallic state.~\cite{ci21} In experiment, the electron diffraction analysis for La$_{1-x}$Ca$_{x}$MnO$_3$, in a narrow region around $x=0.5$, shows us that the charge order (CO) occurs below the Curie temperature, i.e., inside the FM state.~\cite{ci8} In the so-called CO-FM phase, a band gap opens and the system is an insulator.~\cite{ci8} To explain the complexity of the phase structures and also the transport properties in the entire temperature range, the DE model thus needs to be improved. One possible amendment addressing the coexistence of CO and FM states is to involve a large Jahn-Teller lattice distortion coupling to the itinerant electrons. This distortion causes a metal-insulator transition (MIT) via strong polaronic narrowing of the conduction electron band. However, the Jahn-Teller lattice distortion makes the CO phase stabilize above the FM transition.~\cite{ci7} This is not the situation observed in experiments.~\cite{ci2,FFI98,ci8} Alternatively, the interplay between the CO and FM states might also be explained by adding a random local potential to the DE model.~\cite{ci13,PT05} The random local potential comes from the $D^{2+}$ doping, arising from random substitution of $T^{3+}$ by $D^{2+}$. This so-called diagonal-disorder is inevitable in doped manganites.~\cite{ci1,ci2,ci9} In the present paper, we use the key idea that the diagonal-disorder can be modeled by the Falicov-Kimball (FK) model.~\cite{ci10} Although the FK model is simple, it contains a rich variety of phases possibly controlled by the electronic interaction.~\cite{ci11,ci12} Incorporating the FK-type diagonal-disorder into the DE model, one expects that an interplay between the electronic correlation and the disorder might trigger the complex anomalies observed in the systems that the DE model alone fails to interpret. The phase structure observed in the combined model shows us that at low temperature a checkerboard CO state stabilizes below the Curie temperature, i.e., inside the FM state.~\cite{ci13,PT05} However, there is lack of evidence to clarify the existence of an MIT, in particular, its relation with the CO-FM transition as observed in the experiments.~\cite{ci8} In our work, detailed signatures of the electronic density of states (DOS) and transport properties in a large range of temperatures are addressed to discuss the MIT in connection with the CO-FM transition. 

To analyze the electronic DOS of the DE model including the diagonal disorder, we employ a dynamical mean-field theory (DMFT).~\cite{ci14} The DMFT has been extensively used for investigating strongly correlated electron systems.~\cite{ci14} It is based on the fact that the self-energy depends only on the frequency in the infinite dimensional limit. Dividing the system into two sublattices, DMFT permits us to evaluate explicitly the electronic Green functions and then the carrier densities for individual spins on each sublattice site. The magnetization, CO order parameter as well as the transport coefficients in the Kubo-Greenwood formalism therefore are straightforwardly evaluated. The MIT accompanied by the FM-CO transition is then discussed. At extremely low temperatures, it is found that a finite disorder drives the checkerboard CO state inside the FM regime. In the CO state, the electronic resistivity is extremely enhanced while the thermal conductivity is suppressed. Those qualitative features typify an insulating state. Whereas at large temperature the system stabilizes in the metallic state. The FM-CO transition temperature is exactly the MIT temperature.

This paper is organized as follows. In Sec. II, we introduce a microscopic Hamiltonian essentially applied to doped manganites. Dividing the system into two sublattices, transport coefficients calculated to the Hamiltonian in the Kubo-Greenwood formalism are derived. Section III outlines the DMFT application, an explicit single-particle Green's function for an individual sublattice is then calculated. In Sec. IV, we present the numerical results and their discussion. Our conclusions can be found in Sec. V.

\section{Model and transport coefficients}
The DE model combined with the diagonal disorder is proposed in the following Hamiltonian:
\begin{eqnarray}
\mathcal{H}&=&-t\sum_{\langle i,j\rangle,\sigma} c^{\dagger}_{i\sigma} c^{\null}_{j\sigma}
- \mu \sum_{i\sigma}c^{\dagger}_{i\sigma} c^{\null}_{i\sigma}  - 2J\sum_{i} S^{z}_{i} s^{z}_{i}\nonumber \\
&&+U \sum_{i\sigma} n^{}_{i\sigma} n^{f}_{i}+E_{f} \sum_{i} n^{f}_{i},
\label{hamil}
\end{eqnarray}
where $c^{\dagger}_{i\sigma}$($c^{\null}_{i\sigma}$) is the creation (annihilation) operator for an itinerant electron with spin $\sigma$ at lattice site $i$. The first term in the Hamiltonian~(\ref{hamil}) represents the electron hopping between nearest-neighbor sites, $t$ is the hopping integral, and in the limit $d\rightarrow \infty$, it scales with the spatial dimension $d$ as $t=t^\ast/{2\sqrt d}$.~\cite{ci14} In the following, $t^\ast=1$ is taken as the unit of energy.~\cite{ci14} $S^{z}_{i}$ and $s^{z}_{i}=\sum_{\sigma}\sigma c^{\dagger}_{i\sigma}c^{\null}_{i\sigma}/2$ are the $z$ component of the localized magnetic ion spin and of the itinerant electron spin, respectively. The third term in Eq.~\eqref{hamil} therefore illustrates the Ising-type Hund coupling of the local and the itinerant electrons, $J$ is the strength of the coupling. In the last two terms, $n_{i\sigma}=c^{\dagger}_{i\sigma} c^{\null}_{i\sigma}$ is the itinerant electron occupation number operator and $n^f_{i}$ is a classical variable. $n^f_{i}$ takes the value $1$ ($0$) if site $i$ is occupied (non-occupied) by a $D^{2+}$ ion. Note here that a doped manganite is a mixed-valance compound such as T$^{3+}_{1-x}$D$^{2+}_x$Mn$^{3+}_{1-x}$Mn$^{4+}_x$O$^{2-}_3$, the density of unfavorable Mn$^{4+}$ sites therefore is $x$ that reads $x=\sum_i\langle n^f_i\rangle/N$ ($N$ is the number of lattice sites). $U$ is the strength of the local disorder and is mapped onto the difference in the local potential, which splits energetically in favor of Mn$^{3+}$ and Mn$^{4+}$ ions. Note here that the disorder between these two valance states of the manganese ions is not exactly the original disorder generated by doping of D. It looks like a binary alloy disorder. In this work, we model the binary alloy disorder by a variable (i.e., $n^f_i$) of the Falicov-Kimball model. The chemical potential $\mu$ controls the carrier doping, while $E_{f}$ is included to control the fraction of the sites having the additional local potential. The first three terms in the Hamiltonian describe a simplified DE (SDE) model,~\cite{ci21} whereas the last two terms together with the hopping term form the FK model.~\cite{ci10,ci17,ci18} In the framework of DMFT, the individual simplified DE or FK model, and also their combination have been considered in the literature.~\cite{LF01,ci12,ci13,PT05} In particular, by analyzing the charge and spin susceptibilities of the combined model, the CO has been found to exist inside the FM phase.~\cite{ci13,PT05} In the present work, we adopt the DMFT to study the CO transition in competition with the FM state but in a different way. Here, the CO and FM states are characterized by a CO order parameter and the magnetization, respectively. Signatures of the phases are also discussed by analyzing the density of states (DOS) of the itinerant electrons for an individual spin at a different sublattice site. Thus it is convenient to divide the system into two sublattices $A$ and $B$. Assuming that the number of lattice sites are equal in the two sublattices,~\cite{MSF08} the Hamiltonian~\eqref{hamil} can be rewritten in four terms like
\begin{eqnarray}\label{hamilAB}
\mathcal{H}=\mathcal{H}_{AB}+\mathcal{H}_{BA}+\sum_{\eta=\{A,B\}}\mathcal{H}_\eta,
\end{eqnarray}
where
\begin{equation}
\mathcal{H}_{\eta}=U\sum_{i\sigma}n^f_{i\eta}n^{}_{i\eta\sigma}-2J\sum_{i}S^{z}_{i\eta}s^{z}_{i\eta}+E_f\sum_i n^f_{i\eta}
\end{equation}
is the on-site interaction term, whereas
\begin{equation}
\mathcal{H}_{AB}=-t\sum_{i\sigma}a_{i\sigma}^{\dagger}b^{}_{i\sigma} \quad \textrm{and} \quad \mathcal{H}_{BA}=-t\sum_{i\sigma}b_{i\sigma}^{\dagger}a^{}_{i\sigma},
\end{equation}
correspond to the hopping between $A$ and $B$ sublattice sites. $a_{i\sigma}^{\dagger}$ and $a^{}_{i\sigma}$ ($b_{i\sigma}^{\dagger}$ and $b^{}_{i\sigma}$) are the creation and annihilation operators of the itinerant electron with spin $\sigma$ at site $i$, respectively, in sublattice $A$($B$). Here the sum runs over all lattice sites of one sublattice.

To study the transport properties, in our work, we follow the Kubo-Greenwood formalism.~\cite{ci19} The electrical resistivity $\rho$, the thermal conductivity $\kappa $, and the thermopower $S$ are calculated by linear response theory. At temperature $T$, this gives
\begin{eqnarray}
\rho &=&\frac{T}{e^2L^{11}},\label{sig}\\
\kappa &=& \frac{1}{T^2}\left[L^{22}-\frac{(L^{12})^2}{L^{11}}\right],\label{kap}\\
S&=&-\frac{1}{eT}\frac{L^{12}}{L^{11}},\label{pow}
\end{eqnarray}
where $L^{11}, L^{12}$ and $L^{22}$ are the transport coefficients which are determined from the analytic continuation of the respective correlation functions at zero frequency, i.e.,
\begin{eqnarray}
L^{\alpha\beta}=\lim_{\nu \rightarrow 0}T \textrm{Im} \frac{\bar{L}^{\alpha\beta}(i\nu_n\rightarrow \nu+i0^+)}{\nu}.\label{8}
\end{eqnarray}
Here, $\bar{L}^{\alpha\beta}(i\nu_n)$, where $\alpha,\beta=\{1,2\}$, are the current-current correlation functions and $i\nu_n=(2n+1)\pi T$ is the fermionic Matsubara frequency. Explicit expressions of $\bar{L}^{\alpha\beta}(i\nu_n)$ read
\begin{eqnarray}
\bar{L}^{11}(i\nu_n)&=&\int_{0}^{\beta}d\tau
e^{i\nu_n\tau}\langle \mathcal{T}_\tau {\bf j}(\tau){\bf j}(0)\rangle ,\label{bl1}\\
\bar{L}^{12}(i\nu_n)&=&\int_{0}^{\beta}d\tau e^{i\nu_n\tau}\langle
\mathcal{T}_\tau {\bf j}(\tau){\bf j}_Q(0)\rangle ,\label{bl2}\\
\bar{L}^{22}(i\nu_n)&=&\int_{0}^{\beta}d\tau e^{i\nu_n\tau}\langle
\mathcal{T}_\tau {\bf j}_Q(\tau){\bf j}_Q(0)\rangle, \label{bl3}
\end{eqnarray}
where $\beta=1/T$ and $\mathcal{T}_\tau$ is the time-ordering operator. In Eqs.~(\ref{bl1})--(\ref{bl3}), $\bf j$ and ${\bf j}_Q$ are the particle-current and the head-current operators, respectively. A key point therefore in a calculation of the transport coefficients is to evaluate explicitly expressions of the current operators. The particle-current operator in corresponding to the Hamiltonian $\mathcal{H}$ is defined by
\begin{equation}\label{hp}
{\bf j}= i[\mathcal{H},\bf P],
\end{equation}
where ${\bf P}=\sum_{i}{\bf R}_{i}n_i$ is the polarization operator.~\cite{ci19} For the Hamiltonian given in Eq.~\eqref{hamilAB} one obtains an expression of the particle-current operator depending on time
\begin{eqnarray}
{\bf j}(\tau)=\sum_{{\bf q}\sigma}{\bf v}_{\bf q}[a_{\bf q\sigma }^{\dagger}(\tau)b_{\bf q\sigma}(\tau)+b_{\bf q\sigma }^{\dagger}(\tau)a_{\bf q\sigma}(\tau)]. \label{jpar}
\end{eqnarray}
Here we have written the creation and annihilation operators in momentum space by using the Fourier transformations [i.e., $a^\dagger_{{\bf q}\sigma}=\sum_j a^\dagger_{j\sigma} \exp(i{\bf R}_j{\bf q})/N$, for instance]. ${\bf v}_{\bf q}=\bigtriangledown_{\bf q} \varepsilon (\bf q)$ means the velocity and $\varepsilon (\bf q)$ is the dispersion of the non-interacting electrons. The heat-current operator ${\bf j}_{Q}$ is determined by ${\bf j}_{Q}={\bf j}_{E}-\mu{\bf j}$, where ${\bf j}_E$ is the energy current operator which is constructed by taking a commutator of the Hamiltonian with the energy polarization operator $\sum_i{\bf R}_ih_i$ (note here that $\mathcal{H}=\sum_{i}h_i$). Based on the equation of motion technique,~\cite{ci17} the time dependence of the heat-current operator reads
\begin{align}\label{jq}
{\bf j}_{Q}(\tau)=&\lim_{\tau'\rightarrow \tau^{-}}\frac{1}{2}\sum_{\bf
q\sigma}\left(\frac{\partial}{\partial \tau}-\frac{\partial}{\partial \tau'}\right) {\bf v}_{\bf q}
\nonumber \\
&\times [a_{{\bf q}\sigma }^{\dagger}(\tau)b_{{\bf q}\sigma}(\tau')+b_{{\bf q}\sigma }^{\dagger}
(\tau)a_{{\bf q}\sigma}(\tau')].
\end{align}
From the expressions of ${\bf j}(\tau)$ and ${\bf j}_{Q}(\tau)$, respectively, in Eqs.~(\ref{jpar}) and
(\ref{jq}) we can evaluate the current-current correlation functions given in Eqs.~(\ref{bl1})--(\ref{bl3}). From Eq.~\eqref{8}, the transport
coefficients read in a form~\cite{ci21,MSF08,ZBF14}
\begin{align}
L^{\alpha\beta}&=\int d\omega \left(-\frac{\partial f(\omega)}{\partial \omega}\right)\tau(\omega)\omega^{\alpha+\beta-2},\label{Lij}
\end{align}
where
\begin{equation}
\tau(\omega)\!=\!T\sum_\sigma\!\int \! d\varepsilon \rho(\varepsilon)[A_{A,\sigma}(\varepsilon ,\omega)A_{B,\sigma}(\varepsilon ,\omega)+A^2_{AB,\sigma}(\varepsilon ,\omega)],\label{tau}
\end{equation}
plays the role of the exact many-body relaxation time.~\cite{MSF08} $\tau(\omega)$ in Eq.~\eqref{tau} has been written in units of $2\pi\sigma_0$, where $\sigma_0$ is the unit of conductivity, which is defined in Ref.~\onlinecite{PCJ93}. In Eq.~\eqref{tau}, $\rho(\varepsilon)$ is the density of states of non-interacting electrons and $A_{A(B),\sigma}(\varepsilon ,\omega)$ and $A_{AB,\sigma}(\varepsilon ,\omega)$ are spectral functions. In our problem, the spectral functions are evaluated directly from Green's functions of the itinerant electrons defined below in Eqs.~(\ref{Gaa}-\ref{Gab}), i.e., $A_{\eta,\sigma}(\varepsilon ,\omega)=-\textrm{Im}G_{\eta,\sigma}(\varepsilon ,\omega)/\pi$ ($\eta=\{A,B\}$) and $A_{AB,\sigma}(\varepsilon ,\omega)=-\textrm{Im}G_{AB,\sigma}(\varepsilon ,\omega)/\pi$. In the derivation of the expression in Eq.~\eqref{Lij} we have neglected the vertex corrections which actually vanish in the limit of infinite dimensions. In the limit of infinite dimensions, another exact expression of the relaxation time $\tau(\omega)$ for the spinless Falicov-Kimball model can be found in Ref.~\onlinecite{MSF08}.

The expression of the transport coefficients in Eq.~\eqref{Lij} looks similar to the Mott-Thellung noninteracting form.~\cite{CT61} Here, $f(\omega)=1/[\exp(\omega /T)+1]$ is the Fermi-Dirac distribution function. From the analytical expression of $\tau(\omega)$ in Eq.~\eqref{tau}, one realizes that to evaluate the transport coefficients one needs to determine the single-particle Green's functions $G_{\eta,\sigma}(\varepsilon ,\omega)$ and $G_{AB,\sigma}(\varepsilon ,\omega)$. In the next section, we will point out a calculation of the Green's functions in the framework of DMFT applying to the Hamiltonian written in Eq.~\eqref{hamilAB}.

\section{Dynamical mean-field theory}

Our aim in this section is to evaluate the Green's function of the itinerant electrons in the ideas of DMFT. The key point of the DMFT is that, in the infinite dimensional limit, the self-energy of the electrons is local and depends on the frequency only.~\cite{ci14} For a system involving two sublattices $A$ and $B$, the Green's function of the itinerant electrons can be determined via the Dyson equation
\begin{align}
\hat{G}_{\sigma}({\bf k}, i\omega_n)&=
\left(\begin{array}{cc}
				G_{A,\sigma}({\bf k} ,i\omega_n) &   G_{BA,\sigma}({\bf k} ,i\omega_n)\\
				G_{AB,\sigma}({\bf k} ,i\omega_n)	& G_{B,\sigma}({\bf k} ,i\omega_n)\\
						\end{array}               				\right)\nonumber\\
						&=
\left(\begin{array}{cc}
				\xi^A_\sigma(i\omega_n) &   -\varepsilon({\bf k})\\
				-\varepsilon({\bf k})	& \xi^B_\sigma(i\omega_n)\\
						\end{array}               				\right)^{-1},
\label{gmatrix}
\end{align}
where
\begin{equation}
\xi^\eta_\sigma(i\omega_n)=i\omega_n+\mu -\Sigma^{\eta}_{\sigma}(i\omega_n),
\end{equation}
and $\Sigma^{\eta}_{\sigma}(i\omega_n)$ is the self-energy of the itinerant electrons on sublattice $\eta$, which depends on the frequency only. Here we have assumed that each sublattice still can be considered as an infinite dimensional system. By taking the matrix inverse of Eq.~\eqref{gmatrix} one arrives at the momentum or energy dependence of elements of the Green's functions:
\begin{align}
G_{A(B),\sigma}({\bf k} ,i\omega_n)&=\frac{\xi^{B(A)}_\sigma(i\omega_n)}{\xi^{A}_\sigma(i\omega_n)\xi^{B}_\sigma(i\omega_n)-\varepsilon({\bf k})^2},\label{Gaa} \\ 
G_{AB,\sigma}({\bf k} ,i\omega_n)&=G_{BA,\sigma}({\bf k},i\omega_n)\nonumber\\
&=\frac{\varepsilon({\bf k})}{\xi^{A}_\sigma(i\omega_n)\xi^{B}_\sigma(i\omega_n)-\varepsilon({\bf k})^2}. 
\label{Gab}
\end{align}

In the DMFT framework, the self-energies are determined by solving an effective single-site problem. The effective action of our system reads
\begin{align}
&S^\eta_{\small{\textrm{eff}}}[S^z_\eta,n^f_\eta]=-\int d\tau \int
d\tau'\sum_{\sigma}c_{\eta \sigma}^{\dagger}(\tau){\cal
G}_{\eta,\sigma}^{-1}(\tau -\tau')c_{\eta\sigma}(\tau')
\nonumber \\
&-\int d\tau \sum_{\sigma}(JS^z_\eta\sigma -Un^f_{\eta})c_{\eta\sigma}^{\dagger}(\tau)c_{\eta\sigma}(\tau)+E_fn^f_{\eta},
\label{seff}
\end{align}
where $c_{\eta\sigma}(\tau)$ and $c_{\eta\sigma}^{\dagger}(\tau)$ are Grassmann variables, and ${\cal G}_{\eta,\sigma}(\tau -\tau')$ is the Green's function of the effective medium according to sublattice $\eta$. The local Green's function $G_{\eta,\sigma}(i\omega_n)$ of the effective single-site problem satisfies the Dyson equation
\begin{eqnarray}
G_{\eta,\sigma}^{-1}(i\omega_n )={\cal G}_{\eta,\sigma}^{-1}(i\omega_n)-\Sigma^\eta_{\sigma}(i\omega_n).
\label{g01}
\end{eqnarray}
The local Green function is solely determined within the dynamics of the effective single-site impurity embedded in the dynamical mean-field medium,
\begin{eqnarray}
G_{\eta,\sigma}(i\omega_n )=\frac{\delta {\mathcal{Z}^\eta_{\small{\textrm{eff}}}}}{\delta
{\cal G}_{\eta,\sigma}^{-1}(i\omega_n)},
\label{g1}
\end{eqnarray}
where $\mathcal{Z}^\eta_{\small{\textrm{eff}}}$ is the effective partition function for the sublattice $\eta$. Within the effective single-site problem, the partition function can be determined from the effective action as
\begin{eqnarray}
{\cal Z}^\eta_{\small{\textrm{eff}}}=\textrm{Tr}\int
Dc_{\eta \sigma}^{\dagger}Dc_{\eta \sigma}e^{-S^\eta_{\small{\textrm{eff}}}[S^z_\eta,n^f_\eta]},
\label{zeff}
\end{eqnarray}
where the trace is taken over $S^{z}_\eta$ and $n^f_{\eta}$. For the effective action of the problem addressed in Eq.~\eqref{seff}, the partition function can be evaluated straightforwardly. Indeed, there is no coupling between the local spin $S^z_\eta$ and the impurity density $n^f_{\eta}$ dynamics involved in the effective action, and one can take the trace over $S^z_\eta$ and $n^f_{\eta}$ in (\ref{zeff}) independently. This task in fact is only a simple combination of dealing with the DMFT individually to the SDE and FK models. Finally, we obtain
\begin{align}
{\cal Z}^\eta_{\small{\textrm{eff}}}=&4\sum_{m,n^f_{\eta}}\exp \Big\{-E_fn^f_{\eta}\beta\nonumber\\
&+\sum_{n\sigma }\ln \frac{{\cal G}_{\eta,\sigma}^{-1}(i\omega_n)+J\sigma m -Un^f_{\eta}}{i\omega_n}\Big\},\label{zeff1}
\end{align}
where $m$ takes all possible values of the local spin $S^{z}_\eta$ on the $z$ axis, $m=-3/2,-3/2+1,\dots, 3/2$ and $n^f_{\eta}=0, 1$. From Eq.~(\ref{g1}) one derives an explicit expression of the local Green's function
\begin{eqnarray}
G_{\eta,\sigma}(i\omega_n)=\sum_{m,n^f_{\eta}}\frac{w_{m\eta}(n^f_\eta)}{{\cal G}_{\eta,\sigma}^{-1}(i\omega_n)+J\sigma m -Un^f_{\eta}},
\label{g02}
\end{eqnarray}
where
\begin{align}
w_{m\eta}(n^f_\eta)=&\frac{4}{{\cal Z}^\eta_{\small{\textrm{eff}}}}\exp\Big[-E_fn^f_{\eta}\beta\nonumber\\
&+\sum_{n\sigma}\ln \frac{{\cal G}_{\eta,\sigma}^{-1}(i\omega_n)+J\sigma m -Un^f_{\eta}}{i\omega_n} \Big].
\label{w}
\end{align}
The self-consistent condition of the DMFT requires that the local Green's function in Eq.~\eqref{g02} must coincide with the single-site Green's function of the original lattice in Eq.~\eqref{gmatrix}. That means
\begin{equation}
G_{\eta,\sigma}(i\omega_n) = \frac{1}{N} \sum_{\bf k} G_{\eta,\sigma}({\bf k},i\omega_n).
\label{Greenlocal}
\end{equation}
Here we have written the single-site Green's function of the itinerant electrons for an individual sublattice $\eta$. From the expression in Eq.~(\ref{Gaa}) one easily arrives at a frequency dependence only of the single-site Green's function
\begin{equation}
G_{A(B),\sigma}(i\omega_n) =\int d\varepsilon \rho(\varepsilon)\frac{\xi^{B(A)}_\sigma(i\omega_n)}{\xi^A_{\sigma}(i\omega_n)\xi^B_{\sigma}(i\omega_n)-\varepsilon^2},
\label{g03}
\end{equation}
where the summation over momentum {\bf k} in Eq.~\eqref{Greenlocal} has been replaced by an integration over energies weighted by the density of states of non-interacting itinerant electrons, $\rho(\varepsilon)$. In the infinite dimensional hypercubic lattice, it has a Gaussian form, i.e., $\rho(\varepsilon)=\exp(-\varepsilon^2)/\sqrt{\pi}$. Equations~(\ref{g01}), (\ref{g02}), and (\ref{g03}) form a complete set of equations, which self-consistently determines the self-energy and the Green's function of the itinerant electrons.

\section{Numerical results}

\begin{figure*}[htb]
\includegraphics[angle=0,width=0.7\textwidth]{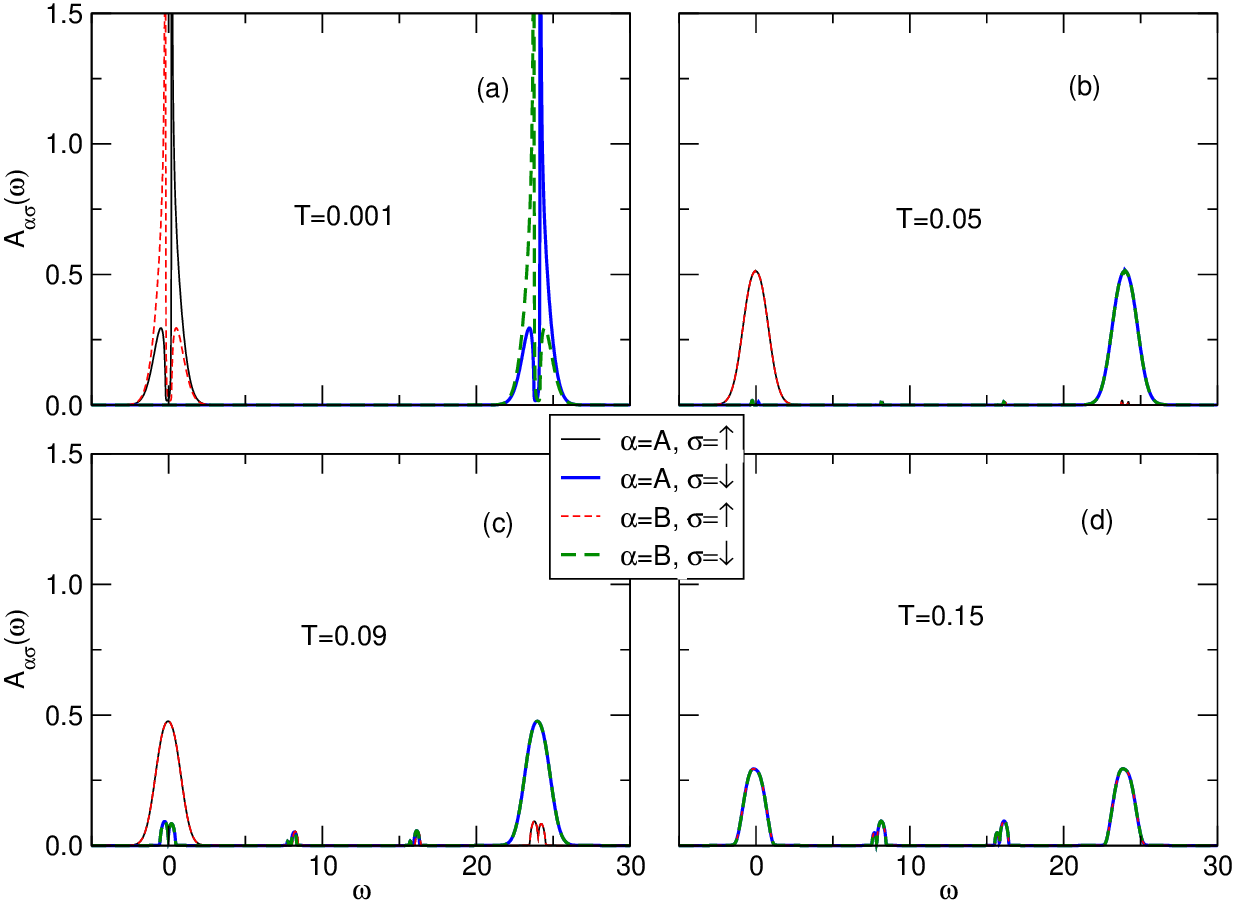}
\caption{(Color online) Density of states $A_{\eta\sigma}(\omega)$ of the itinerant electrons with spin $\sigma$ in sublattice $\eta=A$ and $B$ for $n=0.5$ at different temperatures $T$. Here, we fix $J=8$ and $U=0.4$.}
\label{fig1}
\end{figure*}

In this section, we present numerical solutions of the self-consistent DMFT Eqs.~(\ref{g01}), (\ref{g02}), and (\ref{g03}). In the calculation, we shall use the condition $n+x=1$, where $n=\sum_{i\sigma\eta}\langle n_{i\eta\sigma}\rangle/N$ is the carrier density and $x=\sum_{i\eta}\langle n^f_{i\eta}\rangle/N$. From Eq.~\eqref{w}, one can show that $x=\sum_{m\eta} w_{m\eta}(n^f_\eta=1)$.~\cite{PT05} In doped manganites, the ferromagnetic coupling between the localized magnetic spin and the itinerant electron spin always plays a prerequisite role deciding their physical properties.~\cite{CHD,KEGM} In the following, therefore, we analyze numerical results only for large Hund coupling, i.e., $J\gg 1$. To proceed with the task in the real frequency $\omega$, we use the analytical continuation by replacing $i\omega_n=\omega+i0^+$, all summations of the Masubara frequencies in Eqs.~(\ref{zeff1})--(\ref{w}) thus would be changed to integrals of the real frequency.~\cite{ci21}

First of all, we take a short view of the complex phases by analyzing the DOSs of the itinerant electrons. The DOS with respect to each spin $\sigma$ and sublattice $\eta$ is evaluated straightforwardly from its Green's function following $A_{\eta\sigma}(\omega)=-\textrm{Im}G_{\eta\sigma}(\omega)/\pi$. The carrier density then would be evaluated by taking a summation of DOS magnitudes of all possible states below the Fermi level. The signature of the DOS therefore could be different when the system settles on a different phase. For example, the checkerboard CO state is indicated by a difference of the electronic densities on each sublattice while in the FM state one finds a difference between spin-up and spin-down electronic densities. Figure~\ref{fig1} shows the DOSs of the itinerant electrons for different temperatures at $n=0.5$, $J=8$ and $U=0.4$. At low temperature [$T=0.001$, panel (a)], we see that all four DOSs accounting for each spin up and down electrons on each sublattice $A$ and $B$ are distinguished and as a consequence, the system stabilizes in the CO-FM state.

\begin{figure}[t]
\includegraphics[angle=0,width=0.4\textwidth]{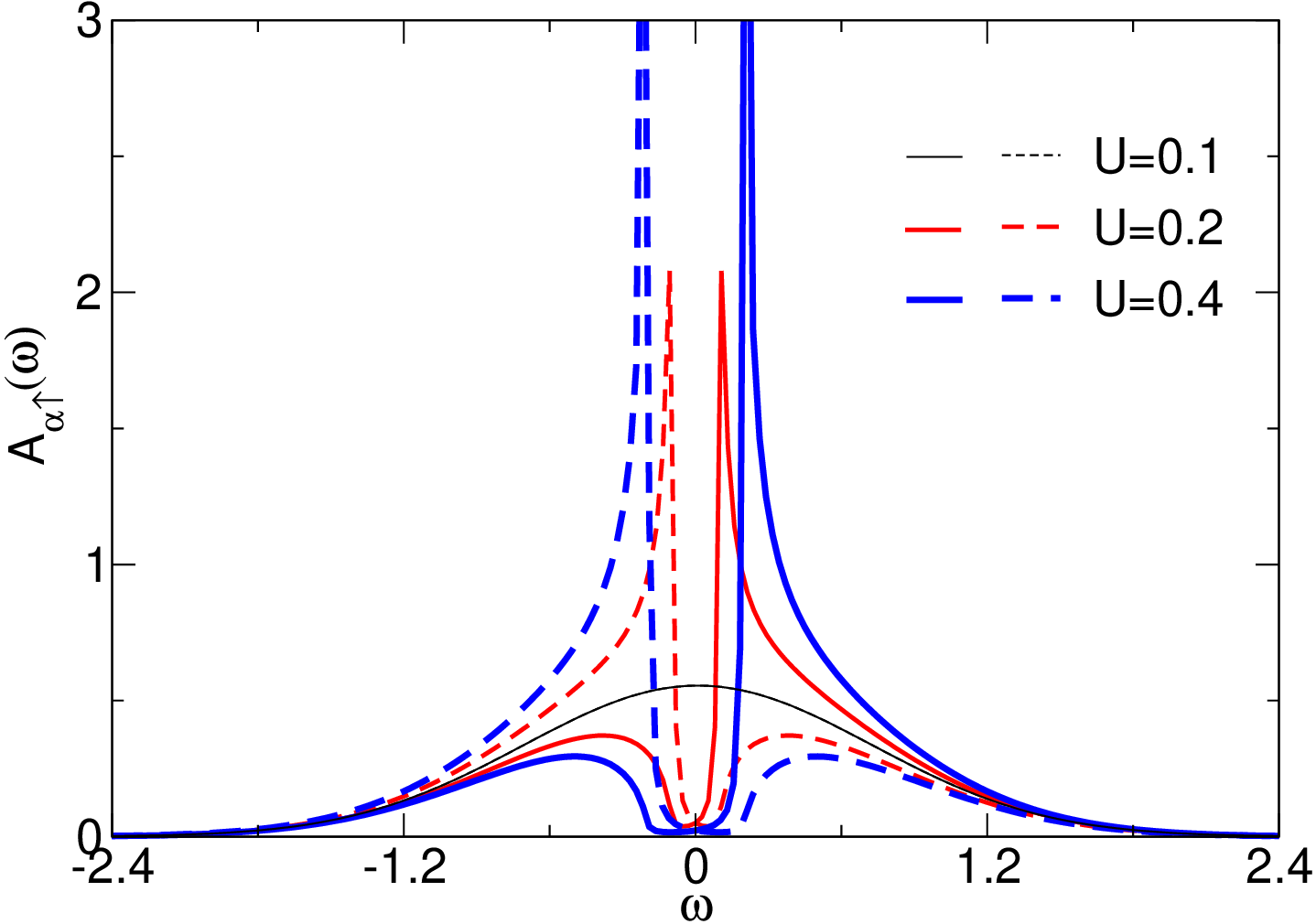}
\caption{(Color online) Density of states of the itinerant spin-up electrons in sublattices $A$ (solid lines) and $B$ (dashed lines) at $n=0.5$ and $T=0.001$ for $J=8$ and different disorders $U$.}
\label{fig2}
\end{figure}

Increasing temperature up to $T=0.04$ [panel (b)] the DOSs on sublattices $A$ and $B$ for a distinct spin totally merge into each other. However, in this situation, we still can see a discrepancy between the DOSs of spin-up and spin-down electrons. This evidence indicates that the system now settles in the charge homogeneous FM state. As a function of temperature, we can conclude that the CO state exists inside the FM state. Enlarging temperature further, DOSs of the spin-up and spin-down carriers start developing an overlap [see panel (c)] and they completely consolidate when the temperature is larger than a critical value. At $T=0.15$ [see panel (d)], for instance, the system stabilizes in the charge homogeneous paramagnetic (PM) state. Inspecting the DOSs at energies close to the Fermi level we also realize that at very low temperature [panel (a)] there is a gap open at the Fermi level. An insulating state therefore is found when the system stabilizes in the CO state. Increasing the temperature [panel (b)--(d)], the gap disappears and the system is in the charge homogeneous-metallic state. 

Discussing the effects of the disorder strength in the association of the CO and the insulating states, in Fig.~\ref{fig2}, we show the DOSs of the spin-up electrons by only inspecting the energies around the Fermi level for some values of $U$ at very low temperature, $T=0.001$. For small $U$, the DOSs of the itinerant electrons on sublattices $A$ and $B$ are identical, the electronic densities on all sites thus are the same or the system is in the charge homogeneous state. Increasing $U$, the two DOSs are separated from each other, this indicates the transition to the CO state. In this case, we see a gap opening at the Fermi level. This marks the insulating state. The size of the gap grows with $U$, thus the insulating state here is driven by the strength of the disorder. Unlike the Mott insulator transition in the hypercubic lattice where we can see a tail in the DOS, in this MIT, the gap opens sharply. The signature of the MIT related to the CO-FM transition will be discussed in detail later.

\begin{figure}[t]
\includegraphics[angle=0,width=0.4\textwidth]{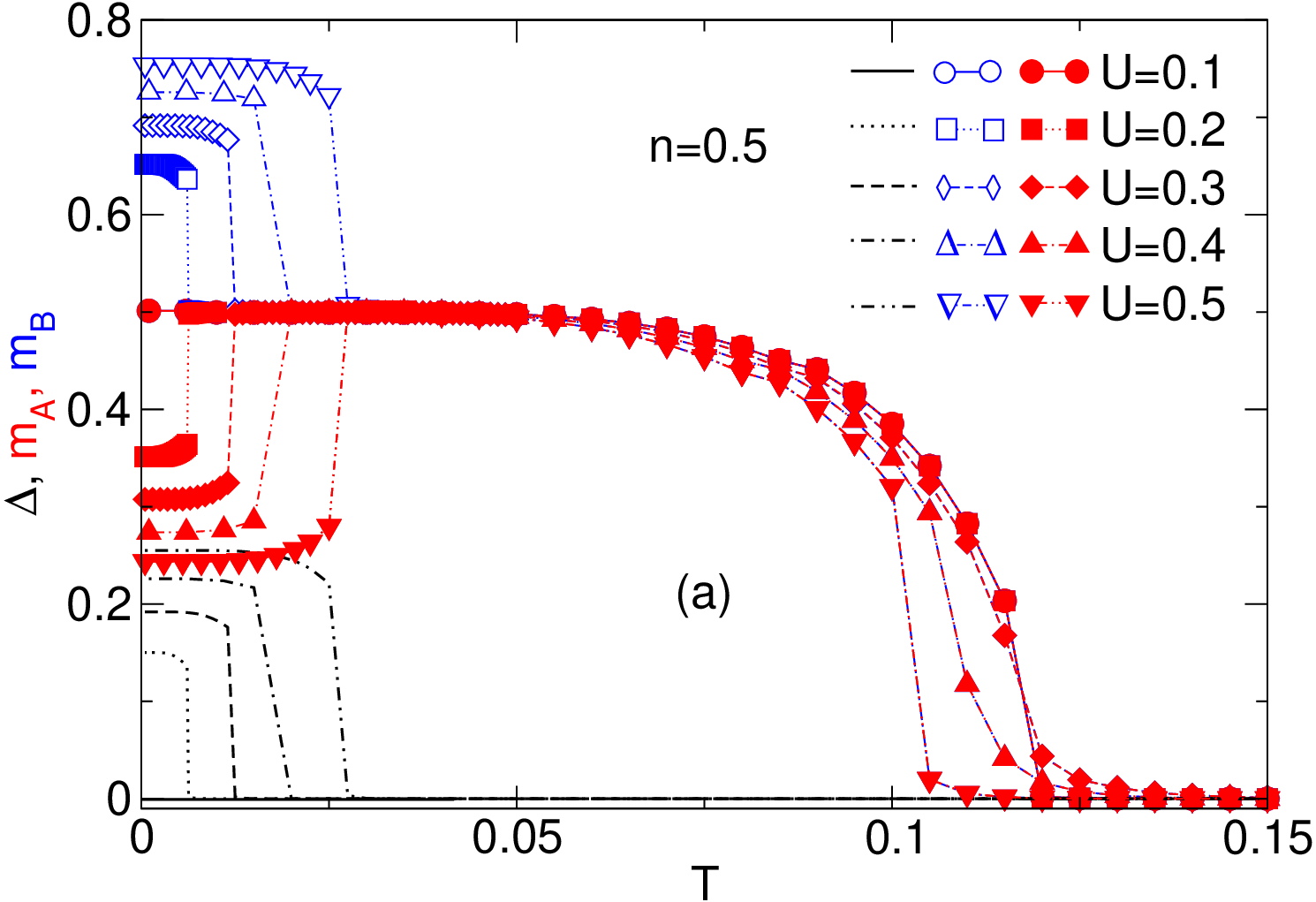}
\includegraphics[angle=0,width=0.4\textwidth]{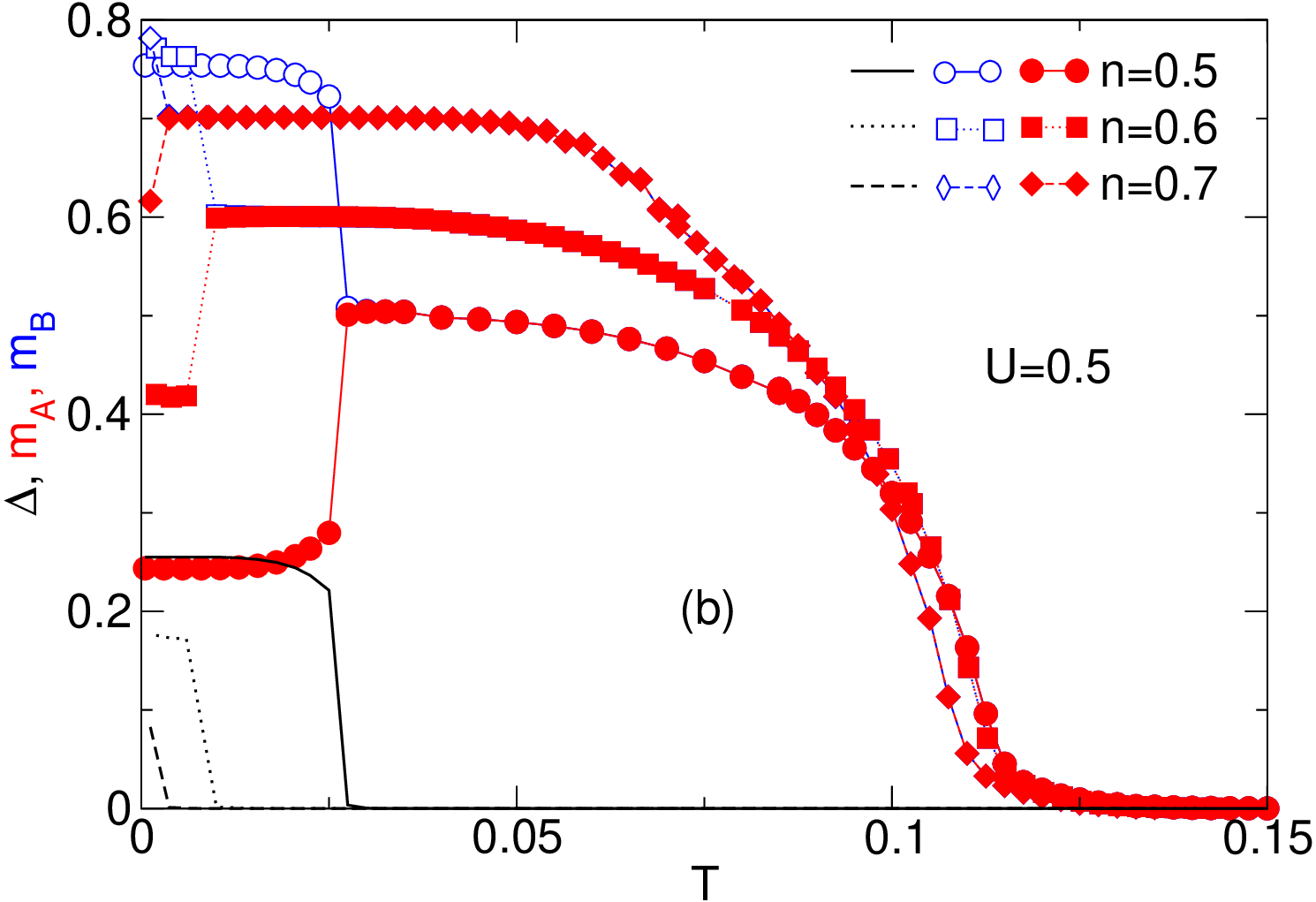}
\caption{(Color online) Dependence of the CO order parameter $\Delta$ (black lines), magnetizations $m_A$ (lines with red filled symbols) and $m_B$ (lines with blue open symbols) on temperature for some values of $U$ at $n=0.5$ (a) and some values of $n$ at $U=0.5$ (b). Here, we have fixed $J=8$.}
\label{fig3}
\end{figure}

To detect the interplay of the CO and the FM states in a more explicit way, we discuss the properties of magnetizations and the CO order parameter. They are respectively defined by
\begin{align}
&m_{\eta}=\frac{1}{N}\sum_{i\in \eta}|\langle n_{i\uparrow}\rangle-\langle n_{i\downarrow}\rangle|,\\
&\Delta=\frac{1}{N}|\sum_{\sigma,i\in A} \langle n_{i\sigma}\rangle-\sum_{\sigma,i\in B}\langle n_{i\sigma}\rangle|,
\end{align}
where $\langle n_{i\sigma}\rangle$ is the density of the spin $\sigma$ itinerant electrons on site $i$. In Fig.~\ref{fig3}(a), we illustrate the magnetization $m_{A(B)}$ of sublattice $A(B)$, and the CO order parameter $\Delta$ as functions of temperature in the case of $n=0.5$ and $J=8$ for different disorders $U$. At small disorder ($U=0.1$) it shows that the CO order parameter is identical to zero in the whole temperature range. In the meanwhile, both magnetizations differ from zero at low temperature and then go to zero when the temperature is larger than a critical value. One can conclude that, if the disorder strength is small, only the charge homogeneous FM state stabilizes at low temperature. At temperatures larger than the critical value, the system is in the charge homogeneous PM state. Increasing $U$, to $U=0.2$ for instance, on the one hand, $\Delta$ is nonzero at low temperature, on the other hand, in this regime, $m_{A}$ and $m_B$ are nonzero but they differ from each other. This point of view once more verifies the existence of the CO state inside the FM phase. At $T>T_{CO}$, the CO order parameter vanishes, $T_{CO}$ therefore indicates a CO transition temperature. In this range of temperatures, one still finds nonzero magnetizations indicating that the system stabilizes in the FM state only. Increasing temperature further depresses the magnetizations and if $T>T_C$ the magnetizations disappear and the system is in the PM state. $T_C$ therefore indicates the FM-PM transition or the Curie temperature. When enhancing the disorder, the CO transition temperature rapidly increases whereas the Curie temperature is slightly depressed. Of course here we have pointed out the result only for small disorder $U$, the behavior of the transition temperatures in a wider range of disorder will be revisited in Fig.~\ref{fig6} below.

The interplay of the CO and the FM states when the itinerant electronic density deviates from $n=0.5$ is illustrated in Fig.~\ref{fig3}(b). Here we have plotted additionally the temperature dependence of $\Delta$, $m_A$ and $m_B$ for $J=8$ and $U=0.5$ at $n=0.6$ and $n=0.7$. In the case of infinite Hund coupling, analyzing the temperature dependence of the static charge and spin susceptibilities indicates that both the CO transition and Curie temperatures reach a maximum at $n=0.5$.~\cite{ci13} In the present case with large Hund coupling, this scenario remains. Moreover, it shows that a slight deviation of the itinerant electronic densities induces significant wiggles of $\Delta$ and therefore in $T_{CO}$, whereas the FM-PM transition temperature $T_C$ is not strongly affected. In the current work, we furthermore indicate that, at low temperature, the magnetizations significantly increase when enhancing the electronic density. This can be explained if we note that increasing the on-site electronic density results in domination of the one type spin density at extremely low temperature, the magnetization thus is built up.

\begin{figure}[htb]
\includegraphics[angle=0,width=0.4\textwidth]{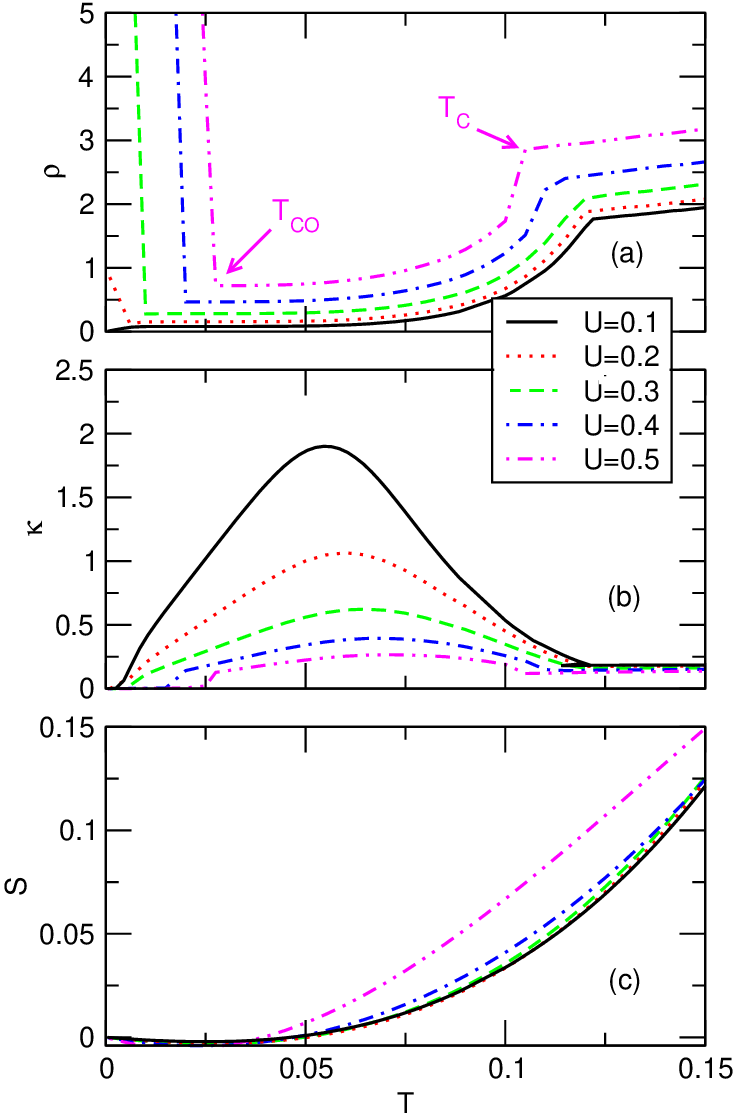}
\caption{(Color online) (a) The electronic resistivity $\rho$, (b) the thermal-conductivity $\kappa$, and (c) the thermopower $S$ as functions of temperature $T$ for some values of $U$ at $n=0.5$ and $J=8$. $T_C$ and $T_{CO}$ respectively indicate the Curie and CO transition temperatures.}
\label{fig4}
\end{figure}

\begin{figure}[htb]
\includegraphics[angle=0,width=0.4\textwidth]{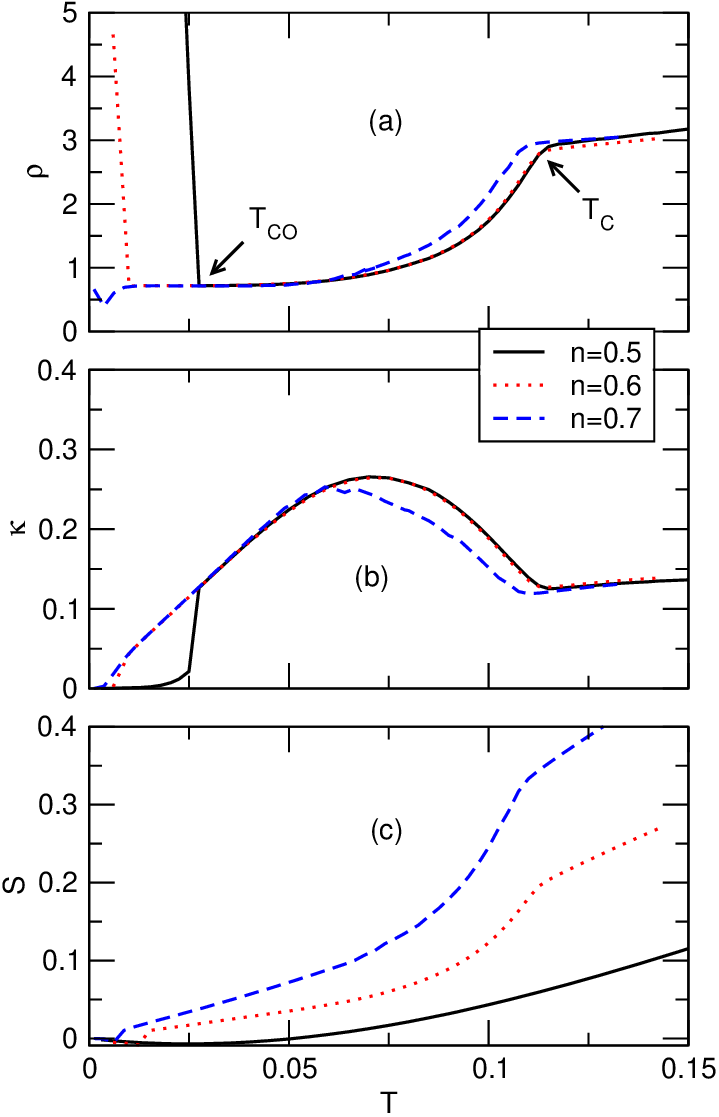}
\caption{(Color online) (a) The electronic resistivity $\rho$, (b) the thermal-conductivity $\kappa$, and (c) the thermopower $S$ as functions of temperature $T$ for some values of $n$ at $U=0.5$ and $J=8$. $T_C$ and $T_{CO}$, respectively, indicate the Curie and CO transition temperatures.}
\label{fig5}
\end{figure}

We start to discuss the transport properties in the system by examining the electronic resistivity depending on the temperature for different disorders at $J=8$ and $n=0.5$ depicted in Fig.~\ref{fig4}(a). Here the electronic resistivity $\rho$ is evaluated following Eq.~\eqref{sig} where the coefficient $L^{11}$ is given in Eq.~\eqref{Lij}. Apparently, the panel shows us that, outside the CO phase, the system settles in the metallic state, indicated by $d\rho/dT>0$. Whereas in the CO state, an opposite situation with $d\rho/dT<0$ happens and the system is an insulator. The $T_{CO}$ here exactly means the MIT temperature. Inside the FM state only, i.e., at temperatures $T_{CO}<T<T_C$, the resistivity declines rapidly by decreasing temperature. That signature can be understood if we note that in the FM state, the magnetic correlation significantly lowers the electronic scattering or enlarges the mean-free path of the itinerant electrons. As addressed in Fig.~\ref{fig2}, we again get a feedback that small disorder does not break the metallic state in the whole temperature range. In this limit, our results can recover those of the SDE model.~\cite{ci21} At large disorder, the DE model with diagonal disorder describes both metallic and insulating states. The kinks at $T_C$ and $T_{CO}$ of the $\rho(T)$ curves respectively indicate the FM-PM and CO transition temperatures. The last two panels (b and c) of Fig.~\ref{fig4} show us the temperature dependence of the thermal conductivity, $\kappa$, and the thermopower, $S$, respectively, for the same parameters given in the panel (a). Lowering the temperature, the thermal conductivity abruptly increases when the system enters the FM state. The enhancement of the thermal conductivity can be understood if one notes that the mean-free path of the conducting electrons is enlarged inside the FM state.~\cite{ci1} Entering the CO state, on the other hand, a gap opens at the Fermi level and the electrons become localized, and as a consequence, the charge and also the heat conductivities are rapidly suppressed. In the whole temperature range, increasing the disorder apparently reduces the electric and also the thermal conductivities. The behavior of the thermal conductivity displayed in Fig.~\ref{fig4}(b) agrees qualitatively with observations for La$_{1-x}$Ca$_x$MnO$_3$ in experiment.~\cite{CNPML97} 

It is well known that the thermopower vanishes whenever there is an electron-hole symmetry [cf. Eq.~\eqref{Lij} with $\alpha=1$ and $\beta=2$]. At extremely low temperature, our case with $n=0.5$ can be considered to be nearly perfect electron-hole symmetric. The high energy spectrum in this case plays a less important role and would be counted out [cf. Fig.~\ref{fig2}]. That results in the thermopower being negligibly small [see Fig.~\ref{fig4}(c)] at low temperature. However, at large temperature, the nearly perfect electron-hole symmetric scenario is invalid, the thermopower thus is nonzero. The thermopower changes its sign in the metallic state.

Next, we continue discussing the transport properties of the system by plotting in Fig.~\ref{fig5} the electronic resistivity, the thermal conductivity, and the thermopower versus temperature for some itinerant electronic densities $n$ at $U=0.5$ and $J=8$. Similar to the behavior of the electronic resistivity discussed in Fig.~\ref{fig4}(a) before, one still finds the MIT at the CO transition temperature [see Fig.~\ref{fig5}(a)]. The MIT temperature decreases if the itinerant electronic density is increased. Temperature dependence of the thermal conductivity and the thermopower shown in Figs.~\ref{fig5}(b) and~\ref{fig5}(c) also provides us a significant difference of thermodynamics signatures between the CO-FM and the FM-only states. At low temperature, the checkerboard CO-FM state stabilizes itinerant electrons, which blocks the hopping of the itinerant electrons between sublattice sites, and therefore suppresses the particle and thus the heat transfers. In the insulating phase, the thermal conductivity becomes exponentially dependent on temperature. That is completely different to the power-law temperature dependence like in the homogeneous charge situation.~\cite{MSF08} In the CO-FM state we find a negligibly small negative thermopower. However, outside the CO state, it changes sign and then strongly monotonically increases with temperature. Increasing the electronic density from $n=0.5$, the chemical potential shifts away from the maximum of the lower band [cf. Fig.~\ref{fig1}(b)], indicating an imbalance of the states below and above around the Fermi level. This gives rise to the thermopower and provides an interpretation of the thermodynamics properties in the ``bad metal".~\cite{ZBF14}

\begin{figure}[htb]
\includegraphics[angle=-0,width=0.4\textwidth]{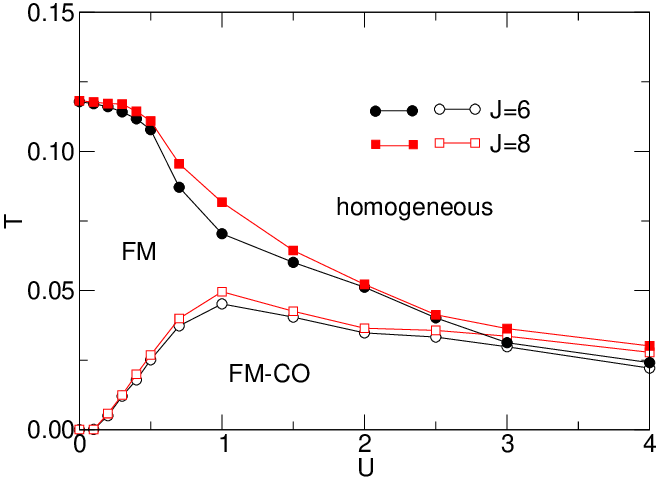}
\caption{(Color online) Curie temperature $T_C$ (filled symbols) and CO transition temperature $T_{CO}$ (opened symbols) depending on $U$ for $J=6$ (circle) and $J=8$ (square) at $n=0.5$.}
\label{fig6}
\end{figure}

Finally, let us summarize the complex structure of the CO and the FM states in the system by plotting a phase diagram in the ($T, U$) plane. Figure~\ref{fig6} presents a disorder $U$ dependence of the transition temperatures $T_{CO}$ and $T_C$ for large given Hund coupling $J$ ($J=6$ and $J=8$) at $n=0.5$. Increasing the disorder, on one hand, increases $T_{CO}$ until it reaches a maximum and then decreases, on the other hand, suppresses $T_C$ in its whole range. At very strong disorder ($U\gg 1$), the two critical temperatures reach to each other but $T_C$ is always above $T_{CO}$, the CO state therefore is only found inside the FM phase. Richer phase diagrams in the model given in Hamiltonian~\eqref{hamil} have been studied by investigating the static charge and spin susceptibilities, the charge ordered and segregated phases coexistence with ferromagnetism depending on doping and disorder have been also discussed.~\cite{PT05} Note here that an investigation for the phase separations directly, as done for the CO and the FM states in this present work, is beyond the scope of this paper.

\section{Conclusion}
To summarize, we have adopted the dynamical mean-field theory to discuss the transport properties in an interplay of the checkerboard charge order and the ferromagnetism states in the double exchange model with a diagonal disorder. By considering the disorder in the form of the Falicov-Kimball model and simplifying the Hund coupling to Ising type, explicit single-particle Green's functions for the itinerant electrons on each sublattice have been found. Magnetizations and the checkerboard charge order parameter therefore are directly evaluated. Following the Greenwood formalism, the transport coefficients have been written in terms of the single-particle spectral functions in the infinite-dimensional limit. This is a simple way to evaluate the electronic resistivity (inverse of the electronic conductivity), the thermal conductivity, and the thermopower. At extremely low temperature, it is found that the checkerboard charge ordering state always exists inside the ferromagnetic regime for finite disorder. In the charge ordering state, the itinerant electrons are blocked and we find a negative derivative of the electronic resistivity with respect to temperature, typifying an insulating state. Whereas, outside the charged ordering phase, in contrast, the system stabilizes in the metallic state. The checkerboard charge ordering-ferromagnetic transition temperature is exactly the MIT temperature. Examining the thermal conductivity, we also find a similar scenario. Indeed, in the checkerboard charge ordering state, the carriers are blocked and then the thermal conductivity is suppressed, whereas it is enhanced in the homogeneous charged state. At extremely low temperature, the thermopower is negligibly small. At large temperature, it changes the sign and then increases, especially if the electronic density deviates from $n=0.5$. That typifies the thermodynamics scenario of the ``bad metal."

\acknowledgements
This research is funded by the Vietnam National Foundation for Science and Technology Development (NAFOSTED) under grant No 103.01-2014.05.

\end{document}